\documentstyle[fleqn,psfig,espcrc2]{article}

\newcommand{\bm}[1]{\mbox{\boldmath$#1$}}
\newcommand{\Done}{\bm{\Delta}^{(\pm)}}
\newcommand{\Dtwo}{\bm{\Delta}^{(2)}}
\newcommand{\LmQCD}{\Lambda_{\mbox{\tiny QCD}}}
\def\lsim{\raise0.3ex\hbox{$<$\kern-0.75em\raise-1.1ex\hbox{$\sim$}}}
\def\gsim{\raise0.3ex\hbox{$>$\kern-0.75em\raise-1.1ex\hbox{$\sim$}}}
\newcommand{\SYMWDIAM}{{\raise-0.4ex\hbox{\LARGE$\diamond$}}}
\newcommand{\SYMWCIRC}{{\raise-0.4ex\hbox{\LARGE$\circ$}}}
\newcommand{\SYMFCIRC}{{\raise-0.4ex\hbox{\LARGE$\bullet$}}}
\newcommand{\SYMWTRIG}{{\raise-0.1ex\hbox{\small$\triangle$}}}

\title{Scaling behavior of $f_{B}$ with NRQCD
       \thanks{Presented by K-I. Ishikawa.}}

\author{JLQCD Collaboration:
  K-I. Ishikawa
   \address{ Department of Physics, Hiroshima
              University, Higashi-Hiroshima, Hiroshima 739-8526, Japan},
  N. Yamada$^{\mbox{\scriptsize\ a}}$,
  S. Aoki
   \address{ Institute of Physics, University of
              Tsukuba, Tsukuba, Ibaraki 305-8571, Japan},
  M. Fukugita
   \address{ Institute for Cosmic Ray Research, 
              University of Tokyo, Tanashi, Tokyo 188-8502, Japan},
  S. Hashimoto
   \address{ High Energy Accelerator Research Organization (KEK),
             Tsukuba, Ibaraki 305-0801, Japan},\\
  N. Ishizuka$^{\mbox{\scriptsize\ b,}\!}$
   \address{ Center for Computational Physics, 
              University of Tsukuba, Tsukuba, Ibaraki 305-8577,
  Japan},
  Y. Iwasaki$^{\mbox{\scriptsize\ b, e}}$,
  K. Kanaya$^{\mbox{\scriptsize\ b, e}}$,
  T. Kaneda$^{\mbox{\scriptsize\ b}}$,
  S. Kaya$^{\mbox{\scriptsize\ d}}$,
  Y. Kuramashi$^{\mbox{\scriptsize\ d}}$,\\
  H. Matsufuru
    \address{ Institut f\"ur Theoretische Physik,
               Universit\"at Heidelberg, Heidelberg, Germany},
  M. Okawa$^{\mbox{\scriptsize\ d}}$,
  T. Onogi$^{\mbox{\scriptsize\ a}}$,
  S. Tominaga$^{\mbox{\scriptsize\ d}}$,
  N. Tsutsui$^{\mbox{\scriptsize\ a}}$,
  A. Ukawa$^{\mbox{\scriptsize\ b, e}}$,
  T. Yoshi\'e$^{\mbox{\scriptsize\ b, e}}$
}

\begin{document}

\begin{abstract}
  We investigate the scaling behavior of the $B$ meson decay
  constant $f_B$ and  $f_{B_{s}}$ at $\beta$$=$$5.7, 5.9, 6.1$,
  employing the NRQCD heavy quark action and the clover light
  quark action.
  Mixing effect from dimension-4 operator in 
  the heavy-light axial-vector current is studied, and we
  find that the $a$ dependence of $f_B$ is significantly
  reduced.
  Our preliminary result for the decay constants in the quenched 
  approximation is 
  $f_{B}$$=$$162(^{+35}_{-18})$  MeV,
  $f_{B_{s}}$$=$$190(^{+40}_{-19})$  MeV,
  and $f_{B_{s}}/f_{B}$$=$$1.18(^{+6}_{-6})$. 
\end{abstract}

\maketitle

\section{Introduction}

A recent development in the NRQCD study of heavy quarks on 
the lattice is the realization that the mixing 
of a dimension-4 operator with the axial-vector current, 
while nominally $O(\alpha_s a)$,  
has a significant effect in the value of the heavy-light 
decay constant\cite{MIXING,GLOK}.   
An investigation of how this mixing effect affects 
the scaling behavior of the decay constant is 
an important issue. 

In this work we study this problem, through simulations, 
with and without the operator mixing taken into account, 
at three values of $\beta$.
A comparison is also made of the present NRQCD results 
with our previous calculation with the relativistic 
heavy quark action\cite{JLQCD}. 

\section{Method}

We describe the light quark by the $O(a)$-improved SW clover 
action with one-loop corrected $c_{\mbox{\tiny sw}}$ as in
Ref.~\cite{JLQCD}.
For heavy quark, we employ two types of the
NRQCD action and operator, one including all terms up to
$O(1/M)$ and the other up to $O(1/M^2)$.  

The $O(1/M)$ NRQCD action we use is
\begin{eqnarray}
&&\!\!\!\!\!\!\!\!\!\!\!\! S\!=\!
\sum_{t,\bm{x}}\!Q(t,\bm{x})\!
\left[ Q(t,\bm{x})-
       \left( 1\!-\!\frac{a H_{0}}{2 n}  \right)^{\!n}\!
       \left( 1\!-\!\frac{a \delta H}{2} \right) 
 \right. \nonumber \\
&&\!\!\!\! \left.\times
     {U^{\dagger}_{4}}\!  
       \left( 1\!-\!\frac{a \delta H}{2} \right)\!\!
       \left( 1\!-\!\frac{a H_{0}}{2 n}  \right)^{\!n}\!
 Q(t-1,\bm{x}) \right], 
\label{eq:H1}
\end{eqnarray}
where $Q$ is a two-component heavy quark field,
$H_{0}$$=$$-\Dtwo$$/$$[2M_{0}]$ and
$\delta H$$=$$ -g$$\bm{\sigma}$$\cdot$$\bm{B}$$/$$[2M_{0}]$.
To the same order in $1/M$, the four-component Dirac field $\psi_{h}$
is related to $Q$ {\it via}
FWT transformation,
\begin{equation}
\psi_{h}(x) = \left(1-\frac{\bm{\gamma}\cdot\Done}{2M_{0}}\right)
\left(
\begin{array}{c}
   Q(x)\\
\chi^{\dag}(x)
\end{array}\right).\label{eqn:FWT}
\end{equation}

The mixing relation between the continuum axial-vector current 
and lattice counterparts, consistently expanded to $O(\alpha_s a)$ 
and $O(\alpha_s/M)$, is given by 
\begin{equation}
A_{4}\!  =\! \left[ 1 
 \!+\! \alpha_{s}\rho^{(0)}_{A} \right]\! J^{(0)}
 \!+\! \alpha_{s}\rho^{(1)}_{A}        \! J^{(1)}
 \!+\! \alpha_{s}\rho^{(2)}_{A}        \! J^{(2)},
\label{eqn:FULLA4}
\end{equation}
where 
$J^{(0)}$$=$$\bar{\psi}_{l}\Gamma \psi_{h}$
with $\psi_l$ the light quark field and $\Gamma$$=$$\gamma_5\gamma_4$, 
$J^{(1)}$$=$$-\bar{\psi}_{l}\Gamma \bm{\gamma}$$\cdot$$a\Done \psi_{h}$
and 
$J^{(2)}$$=$$\bar{\psi}_{l}$
$\bm{\gamma}$$\cdot$$\stackrel{\leftarrow}
{a\bm{\Delta}}$$^{(\pm)}\Gamma \psi_{h}$. 
An important point observed in the first calculation of the 
one-loop coefficients $\rho_A^{(0,1,2)}$\cite{MIXING} 
is that the coefficient $\rho_{A}^{(2)}$ is not suppressed
by $1/aM$ and remains as $O(1)$ for heavy quark, so that
the mixing with the $J^{(2)}$ operator yields a large
$O(\alpha_{s}a\LmQCD)$ contribution.
We have calculated the mixing coefficients for our $O(1/M)$ NRQCD action
which is slightly different from that of Ref.~\cite{MIXING}. 

\begin{table}[t]
\caption{Parameters of simulation.}
\label{tab:PARA}
\begin{tabular}{c|ccc}\hline
$\beta$& 6.1 & 5.9 & 5.7 \\ \hline
Vol.  &$24^{3}\times 64$&$16^{3}\times 48$&$12^{3}\times 32$ \\
\# of conf.&120&300&300 \\
$a^{-1}$[GeV]& 2.29 & 1.60 & 1.04 \\ \hline
\end{tabular}
\vspace{-2.0em}
\end{table}

\section{Results on mixing effects}

We carry out simulation at three values of $\beta$ employing lattices 
and statistics as listed in Table \ref{tab:PARA}. 
To set the lattice scale, we interpolate string tension data 
collected in Ref.~\cite{STR} and set $\sqrt{\sigma}$$=$$427$ MeV. 

Figure \ref{fig:PHIFULVSMUL} shows
our results for the quantity $\Phi_{P}$
$=$$(\alpha_{s}(M_P)$$/$$\alpha_{s}(M_B))^{2/\beta_{0}}$$
f_{P}\sqrt{M_P}$ at $\beta$=5.9.
We observe that
the contribution of the mixing operators ($O(\alpha_{s}a)$),
which is the difference between 
 (\SYMWCIRC's) and (\SYMFCIRC's) in the figure, 
is as large as that of the multiplicative renormalization
of the leading operator ($O(\alpha_{s})$), 
which is the difference between (\SYMWDIAM's) and (\SYMWCIRC's).
This effect becomes more significant towards heavier quark
mass due to a large value of $\rho^{(2)}_{A}$ and that of the matrix element 
of $J^{(2)}$, so that the slope of $\Phi_P$ becomes reduced with the
inclusion of the mixing, as observed in Ref.~\cite{GLOK}.
We find this behavior to be more pronounced at $\beta=5.7$.

\newcommand{\figwidth}{3.2in}
\newcommand{\figright}{-0.5em}
\newcommand{\figbottom}{-2.0em}
\newcommand{\figtop}{-3.0em}
\begin{figure}[t]
\vspace*{\figtop}
\begin{center}
\leavevmode\hspace*{\figright}
\psfig{file=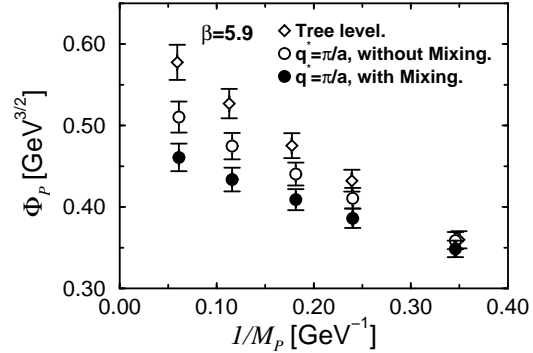,width=\figwidth,angle=-90}
\end{center}
\vspace{-4.0em}
\caption{$1/M_{\mbox{\scriptsize P}}$ dependence of
$\Phi_{\mbox{\scriptsize P}}$ with the $O(1/M)$ NRQCD action
at $\beta=5.9$.}
\label{fig:PHIFULVSMUL}
\vspace{\figbottom}
\vspace*{-2em}
\end{figure}

\begin{figure}[t]
\vspace*{\figtop}
\begin{center}
\leavevmode\hspace*{\figright}
\psfig{file=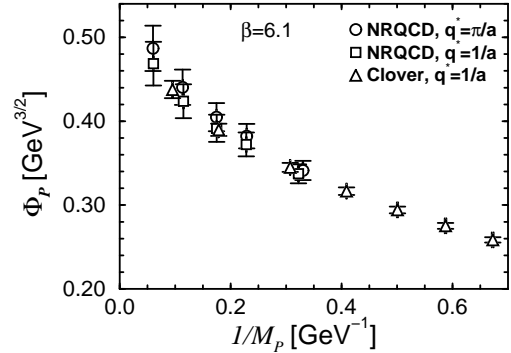,width=\figwidth,angle=-90}
\end{center}
\vspace{-4.0em}
\caption{$\Phi_{\mbox{\scriptsize P}}$ with the $O(1/M)$ NRQCD action
 (this work) and with the SW clover action for heavy quark\cite{JLQCD} at
  $\beta=6.1$.  Mixing effects are not included for both results.} 
\label{fig:PHINRVSCL}
\vspace{\figbottom}
\end{figure}

In Fig.~\ref{fig:PHINRVSCL} we compare 
results for $O(1/M)$ NRQCD action with previous JLQCD results\cite{JLQCD} 
obtained with the SW clover action for heavy quark, 
interpreted as a non-relativistic action within the 
Fermilab formalism\cite{EKM}, at $\beta$$=$$6.1$.
Since the latter calculation does not 
include the effect of $O(\alpha_s a)$ mixing, 
we plot NRQCD results for the one-loop corrected leading operator. 
A good agreement of results for the two actions provides 
a check of viability of both the $1/M$ expansion approach of NRQCD and 
the Fermilab interpretation of the clover action for heavy quark.

\begin{figure}[t]
\vspace*{\figtop}
\begin{center}
\leavevmode\hspace*{\figright}
\psfig{file=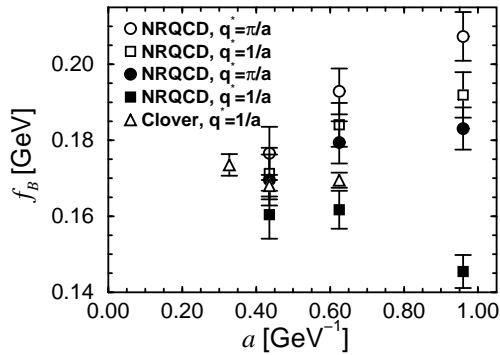,width=\figwidth,angle=-90}
\end{center}
\vspace{-4.0em}
\caption{Scaling behavior of $f_B$ with the $O(1/M)$ NRQCD action. 
Results with the SW clover action for heavy quark\cite{JLQCD}
are also plotted.  
}
\label{fig:AVSFB}
\vspace{\figbottom}
\end{figure}

Figure \ref{fig:AVSFB} presents the scaling behavior of
$f_B$ without (open symbols) and with (filled symboles) operator mixing, 
and for two choices of the momentum scale $q^*$$=$$\pi/a$ and 
$1/a$ for the coupling constant. 
A large scatter of the values at $a^{-1}$$\approx$$1$
GeV$^{-1}$ ($\beta$$=$$5.7$) 
shows that one-loop estimates of renormalization factors are 
not reliable at such a large lattice spacing. 
This  problem is substantially alleviated at 
$\beta$$=$$5.9$ and 6.1 ($0.4\lsim a\lsim 0.6$ GeV$^{-1}$). 
In this region, the NRQCD result without including the operator 
mixing contribution has a large $a$ dependence, which is sizably 
reduced with full inclusion of the mixing. 

It is gratifying that the value of $f_B$ in this range of $a$ 
are reasonably consistent with the results from the clover 
quark action (\SYMWTRIG's)\cite{JLQCD} over the same range. 
Strictly speaking, such a comparison is to be made with the 
continuum extrapolated value of the latter.  
A mild scaling violation exhibited by the clover result suggests
that the agreement would not be severely violated in such an 
extrapolation. 
Two points, however, have to be checked to consolidate the agreement:
(i) the NRQCD values suffer from $O(\alpha_{s}/(aM)^{2})$ errors toward 
smaller lattice spacing, whose magnitude in our range 
of $a$ needs to be examined.  (ii) The clover result does not 
incorporate effects of the $O(\alpha_s a)$ mixing, 
whose magnitude is yet unknown. 

\section{Results for decay constants}

We estimate the physical value of the heavy-light decay 
constants from results at $\beta$$=$$6.1$ obtained with the 
$O(1/M)$ NRQCD action. Since the value of $q^*$ is not known, we take 
the static result $q^*$$=$$2.18/a$~\cite{STATIC} as a guide, 
and calculate the central value from an average of results 
for $q^*$$=$$\pi/a$ and $1/a$.
We then find that
\begin{eqnarray}
f_{B}    \!\!\!&=&\!\!\!
       162(7)(5)(5)(11)(6)(^{+31}_{-8})\mbox{\ MeV},\\
f_{B_{s}}\!\!\!&=&\!\!\!
       190(5)(5)(5)(13)(6)(^{+39}_{-9})(^{+4}_{-0})\mbox{\ MeV}. 
\end{eqnarray}
The first error is statistical including that from chiral
extrapolation.  Remaining are systematic errors arising from 
(i)   the uncertainty of $q^{*}$ estimated by dispersion of results 
      for $q^*$$=$$\pi/a$ and $1/a$, 
(ii)  $O(1/M^{2})$ corrections estimated from comparison of results 
      with the $O(1/M)$ and the $O($$1$$/$$M^2$$)$ calculations,
(iii) $O(\alpha_{s}$$/$$(a$$M)^{2})$ errors estimated
      by dividing $O(\alpha_{s}/(aM))$ contribution, which is
      derived from the result with static perturbative correction, by $a$$M$,
(iv)  scaling violation from comparison of the values at $\beta$$=$$6.1$
      with those at $\beta$$=$$5.9$ and $5.7$, 
and 
(v)  uncertainty in $a^{-1}$, where the upper and lower errors
     correspond to the choice 
     $a^{-1}$$=$2.62 GeV from charmonium 1s-1p splitting and 
     2.21 GeV from $f_K$ as quoted in 
     Ref.~\cite{FNAL}, respectively.
For $f_{B_{s}}$ the central value is obtained with 
$\kappa_s$ for strange quark fixed by $m_K$, and
the last error is estimated from the shift when $\kappa_s$ is 
derived from $m_\phi$.
An $O($$\alpha_{s}$$\LmQCD$$/$$M)$ 
error coming from the action (\ref{eq:H1}) is not included.
A na\"{\i}ve estimate of this error gives $\sim$2\%
at $\beta$$=$$6.1$.

Some systematic errors cancel in the ratio
\begin{equation}
f_{B_{s}}/f_{B}=1.18(3)(5)(^{+2}_{-0}),
\end{equation}
where the statistical error, scaling violation, and the uncertainty of
$\kappa_{s}$, which remain, are given in this order.



\vspace*{3mm}
This work is supported by the Supercomputer Project No.32\,(FY1998)
of High Energy Accelerator Research
Organization\,(KEK), and also in part by the Grants-in-Aid
of the Ministry of Education (Nos.~08640404, 09304029, 10640246,
10640248, 10740107, 10740125).
S.K., H.M. and S.T. are supported by the JSPS Research Fellowship.


\end{document}